\documentclass[%
 reprint,
superscriptaddress,
 amsmath,amssymb,
 aps,prx,
]{revtex4-2}

\usepackage{xcolor}
\usepackage{graphicx}
\usepackage{dcolumn}
\usepackage{bm}
\usepackage{xr}

\makeatletter

\begin{document}

\preprint{APS/123-QED}

\title{Self-organization to multicriticality}

\author{Silja Sormunen}
\affiliation{Department of Computer Science, Aalto University, 00076 Espoo, Finland
}%
\author{Thilo Gross}
\affiliation{
Helmholtz Institute for Functional Marine Biodiversity (HIFMB), 26129 Oldenburg, Germany} 
\affiliation{Alfred-Wegener Institute, Helmholtz Centre for Marine and Polar Research,  27570 Bremerhaven, Germany}
\affiliation{Institute for Chemistry and Biology of the Marine Environment (ICBM), Carl-von-Ossietzky University, 26129 Oldenburg, Germany
}
\author{Jari Saram{\"a}ki}%
\affiliation{Department of Computer Science, Aalto University, 00076 Espoo, Finland
}%

\begin{abstract}
Self-organized criticality has been extensively studied in dynamical systems and, more recently, in networks. However, prior work has focused on organization to a single phase transition controlled by one parameter. Here, we present the first demonstration of self-organization to multicriticality, in which a system tunes itself to the onset of two distinct but simultaneous transitions. We study a model of coupled oscillators on a random network, where the network topology evolves in response to the oscillator dynamics. We show that the system first self-organizes to the onset of oscillations and then drifts to the onset of pattern formation while remaining at the oscillatory threshold, thus becoming critical in two different ways at once. The observed evolution to multicriticality is robust generic behavior that we expect to be widespread in self-organizing systems. Overall, these results reveal the rich dynamical possibilities within self-organized critical states and offer a unifying framework for studying systems, such as the brain, where multiple phase transitions may be relevant for proper functioning.



\end{abstract}

\maketitle

\section{Introduction}
Traditionally, self-organized criticality has been understood as a process in which a system tunes itself to a critical point separating two phases \cite{Bak, Turcotte,Munoz}. However, when the system in question is an adaptive network, where dynamics and structure coevolve, this behaviour can be more complex \cite{gross2021not}. Recently, we showed that simple adaptive mechanisms can drive a self-organizing system to drift along a critical manifold instead of immediately settling at a static critical point \cite{critical_drift}. During this drift, the system's structure continues to evolve while the system remains at criticality. Based on these results, we hypothesized that if the critical manifolds associated with two distinct phase transitions intersect, a system might drift to such an intersection and thus become critical in multiple different ways at once. In this paper, we confirm this hypothesis by demonstrating for the first time the phenomenon of self-organized multicriticality, revealing that the dynamics within the self-organized critical state are far richer than previously recognized.

This discovery is relevant for all adaptive systems hypothesized to self-organize to criticality, including, as a prominent example, the brain. According to the critical brain hypothesis, the brain self-organizes to a critical state driven by adaptive mechanisms that shape the connections between neurons \cite{Chialvo, Herz, pearlmutter2009hypothesis, Hesse, Rybarsch, Timme, Zeraati}. This critical state is thought to provide optimal computational properties, such as optimal information transmission  \cite{Beggs2003} and maximal capacity for information storage \cite{Haldeman}. Although the hypothesis has gained both theoretical and empirical support \cite{Bornholdt2003, MeiselModel, Levina2007, Droste, Yaghoubi, Habibollahi, MeiselEcog, Kitzbichler, Hansen}, there is no consensus on the nature of critical dynamics involved \cite{Wilting2014years}. Some researchers study the onset of synchrony \cite{Safaeesirat}, while others focus on the onset of activity \cite{Kossio} or the onset of chaos \cite{Toker}. It has also been proposed that different phase transitions may play distinct roles across tasks or developmental stages \cite{kanders, Kanders2020}, but a theoretical understanding of how these transitions relate to each other has been lacking. Our results address this gap, showing that a system can move to a new critical regime while staying poised at another.

To demonstrate self-organized multicriticality, we study a model of coupled oscillators on a weighted random network, where edge weights evolve in response to local system dynamics without any centralized control. We show that the system first self-organizes to the onset of oscillations, and then drifts along the critical manifold to multicriticality, where a Turing bifurcation occurs and patterns begin to emerge. Using language from both the theory of critical phenomena and dynamical systems theory, we use the term multicriticality to refer to a system sitting at two or more bifurcations at once, or, phrased differently, at a bifurcation of higher codimension. We show that even after reaching multicriticality, the network topology continues to evolve. Together, these findings provide a unifying framework for understanding systems whose functioning might require different bifurcations or their co-occurrence depending on the environment or task at hand.

\section{Model}

\subsection{Dynamics at nodes}
We study a system of coupled oscillators consisting of an undirected network of $N$ nodes, where the state of each node evolves according to the FitzHugh-Nagumo equations. FitzHugh-Nagumo model \cite{Fitzhugh,Nagumo} is a simplified and analytically tractable version of the Hodgkin-Huxley model describing neuronal excitation. The model equations govern the evolution of two variables $U_i$ and $V_i$, which represent the state of neuron $i$. The strength of coupling is determined by a real-valued coupling matrix $\mathbf{C}$. The adimensionalised differential equations \cite{gambino} with coupling are given by 
\begin{align}
    \dot{\mathbf{U}} &= -\mathbf{U}^3 + \mathbf{U} - \mathbf{V} - C_{00} \mathbf{L} \mathbf{U} - C_{01} \mathbf{L} \mathbf{V}\\
    \dot{\mathbf{V}} &= b (\mathbf{U}-a \mathbf{V}) - C_{10} \mathbf{L} \mathbf{U} - C_{11} \mathbf{L} \mathbf{V},
    \label{diff_eqs}
\end{align}
\noindent where $a$ and $b$ are positive constants and $\mathbf{L}$ is the network Laplacian. Denoting the adjacency matrix of a network with $A$, the Laplacian is constructed by setting $ L_{ii} = \sum_j A_{ij}$ for the diagonal elements, and $L_{ij} = -A_{ij}$ for the off-diagonal elements. Note that our aim in this work is not to model biological networks of neurons but rather to demonstrate with a general model of coupled oscillators the phenomenon of self-organized multicriticality. Consequently, we do not restrict the parameters in Eqs.~(1)-(2) to biologically realistic values and allow the coupling coefficients to take negative values. 

The model exhibits a wide range of dynamical behaviors. Depending on the chosen parameters, the system can undergo a Hopf bifurcation, where the steady-state values of $U$ and $V$ start to oscillate, or a Turing bifurcation, where these values start to differ across nodes. As shown in Refs.~\cite{Segel, Pecora, Nakao, cm2}, the master stability function approach enables us to separate the effects of node dynamics and network topology on the overall dynamics. 


As usual in dynamical systems theory, the eigenvalues of the Jacobian of the system give insight into the model's dynamical behavior. 
As shown in Ref.~\cite{cm2}, the eigenvalues $\mu_i$ of the Jacobian of the whole system can be computed with the equation
\begin{equation}
    \text{Ev}(\mathbf{J}) = \bigcup_{l=1}^{L} \text{Ev} (\mathbf{J}_l) = \bigcup_{l=1}^{L} \text{Ev} (\mathbf{P}-\lambda_l \mathbf{C}),
\end{equation}
\noindent where $\mathbf{P}$ denotes the Jacobian of the FitzHugh-Nagumo equations without the coupling terms, and $\lambda_l$ denotes the $l$th eigenvalue of the network Laplacian. The system undergoes a Hopf bifurcation when a pair of complex conjugate eigenvalues of $\mathbf{J}$ crosses the imaginary axis, which can be detected by analyzing the trace of the reduced matrix $\mathbf{J}_l$. The bifurcation occurs when the trace of $\mathbf{J}_l$ for any eigenvalue $\lambda_l$ vanishes while the determinant of $\mathbf{J}_l$  stays positive, provided that some general non-degeneracy conditions are satisfied. 

Similarly, the Turing bifurcation from the homogeneous steady state occurs when a real eigenvalue $\mu$ crosses zero. This happens when the determinant for $\mathbf{J}_l$ for any Laplacian eigenvalue $\lambda_l$ crosses from positive to negative. Altogether, these conditions indicate that the Hopf and Turing bifurcations can occur independently and possibly simultaneously depending on the dynamical parameters, coupling strength and the network topology.




\subsection{Network plasticity}
As our goal is to study the joint effect of adaptive mechanisms that simultaneously reshape the network structure, we implement two plasticity mechanisms that modify edge weights. Each mechanism acts locally: each node modifies the weights of its links based only on information about its own state and the states of its neighbors. These mechanisms follow a general principle for self-organization \cite{Droste}, which requires a slow process pushing the system towards supercriticality at a constant speed, and a fast process pushing the system towards subcriticality when a node detects that its state is indicative of supercriticality. We call the first the supercriticality rule and the second the subcriticality rule.  

In isolation, the two rules of the first mechanism are expected to push the system towards the onset of oscillation. With the master stability function approach, we can plot the trace of $\mathbf{J}_l$ as a function of the Laplacian eigenvalues $\lambda_l$; if the dynamical parameters  and coupling coefficients in Eq.~\ref{diff_eqs} are chosen so that this function is increasing, the oscillations are born when the trace of $\mathbf{J_1}$ for the leading eigenvalue $\lambda_1$ crosses from negative to positive at a critical value $\lambda_H^*$. To slowly move the system towards supercriticality, each node increases its edge weights that are at least as large as the node's median edge weight by amount $\delta w$ at a constant rate, effectively increasing the leading Laplacian eigenvalue $\lambda_1$. To move toward subcriticality, a node decreases its edge weights at least as large as the median by amount $\delta w$ if the maximum amplitude of oscillations in the node's time series of $U/V$ during the last $t$ time steps exceeds a threshold $\theta_{H}$. The edge weights are not allowed to become smaller than $w_{\min}$.


At the Turing bifurcation, the steady-state values of $U_i$ and $V_i$, denoted by $U_s$ and $V_s$, start to differ between nodes. This occurs when the determinant of the system Jacobian crosses from positive to negative. The master stability function approach shows that the stability is lost when the reduced Jacobian $\mathbf{J}_l$ for any of the Laplacian eigenvalues first becomes negative. We choose the parameters of Eq.~\ref{diff_eqs} so that the function for the reduced determinant is an upward-opening parabola as a function of $\lambda_l$ (see SI II for illustration),  setting $a=0.8, b = 10.5$ and 
\begin{equation}
   \mathbf{C} = \begin{bmatrix}
-1.4 & 0.3\\
-6.8 & 0.9
\end{bmatrix}.
\end{equation}
If the second smallest Laplacian eigenvalue increases to the point that it exits the region where the corresponding determinant is negative, the determinant of the system Jacobian must be positive indicating that the system is in subcritical state (note that the smallest Laplacian eigenvalue is always zero). Hence, self-organization rules modifying the second smallest Laplacian eigenvalue have the potential to drive the system to the Turing bifurcation.

The second smallest eigenvalue is known to reflect how well the network is connected in general (see \emph{e.g.}~\cite{Nair}), and so it can be effectively modified by reshaping parts of the network that have lower-than-average connectivity. Hence, to push the system towards supercriticality, each node decreases the weights of all edges that are weaker than or equal to the node's median edge weight at a constant rate by amount $\delta w$. The opposite rule pushing the system towards subcriticality is activated if a node's average value of $U/V$ during the last $t$ time steps differs more than threshold $\theta_T$ from its neighbors' states. In this case, edge weights smaller or equal than the median are increased by amount $\beta \delta w$, where $\beta$ ensures a time-scale separation between the Turing and Hopf subcriticality rules.

We integrate the model with the explicit Runge-Kutta method of order 4th with 5th order error estimation.
In practice, we perform the topology updates in discrete rounds by first integrating the differential equations for $s$ time units, after which each node checks whether it violates the Hopf or the Turing conditions and updates its edge weights accordingly. To ensure time scale separation between super- and subcriticality updates, each node performs the supercriticality updates with probability 0.5 on every $c$th update round. Before starting the next integration, we add random uniform noise between -0.01 and 0.01 to all values of $U$ and $V$ to ensure that the system does not accidentally get stuck at an unstable equilibrium point.

\begin{figure}
    \includegraphics[width=8cm]{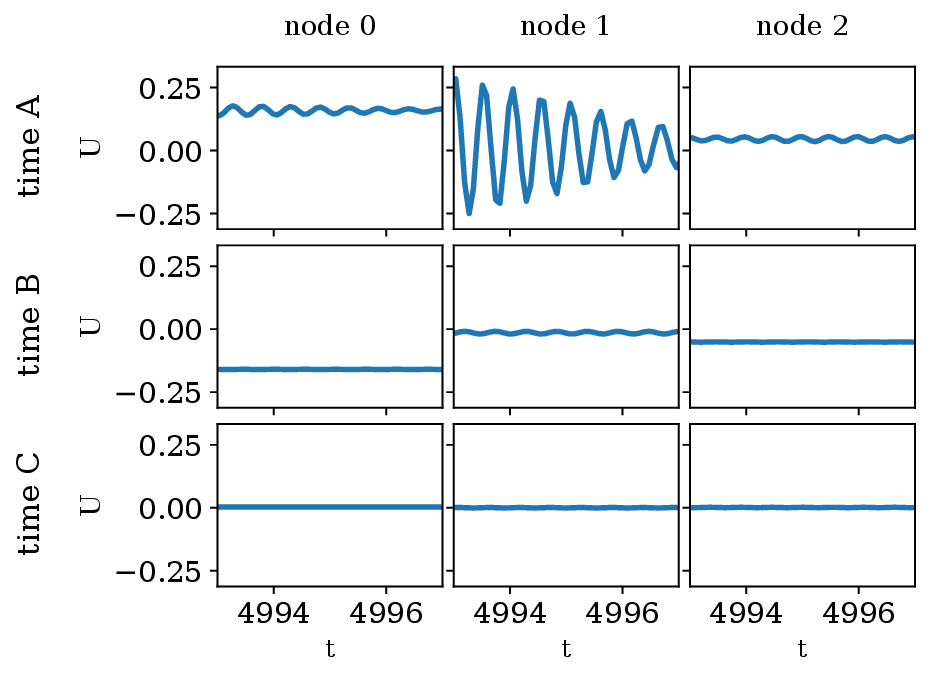}
    \caption{Snapshots of three neighboring nodes' timeseries of variable $U$ at times where neither of the Laplacian eigenvalues $\lambda_1$ and $\lambda_{n-1}$ (time A), only $\lambda_1$  (time B) or both $\lambda_1$ and $\lambda_{n-1}$ (time C) have reached their critical values (time points are indicated in Fig.~\ref{fig:timeseries}a). At time A, the system is supercritical with respect to both Hopf and Turing bifurcation. At time B, the oscillations have died out while there is still variation between the average values of $U$. At time C, the nodes share the same average steady-state value of $U$.}
    \label{3by3}
\end{figure}

\begin{figure*}
    \includegraphics[width=16cm]{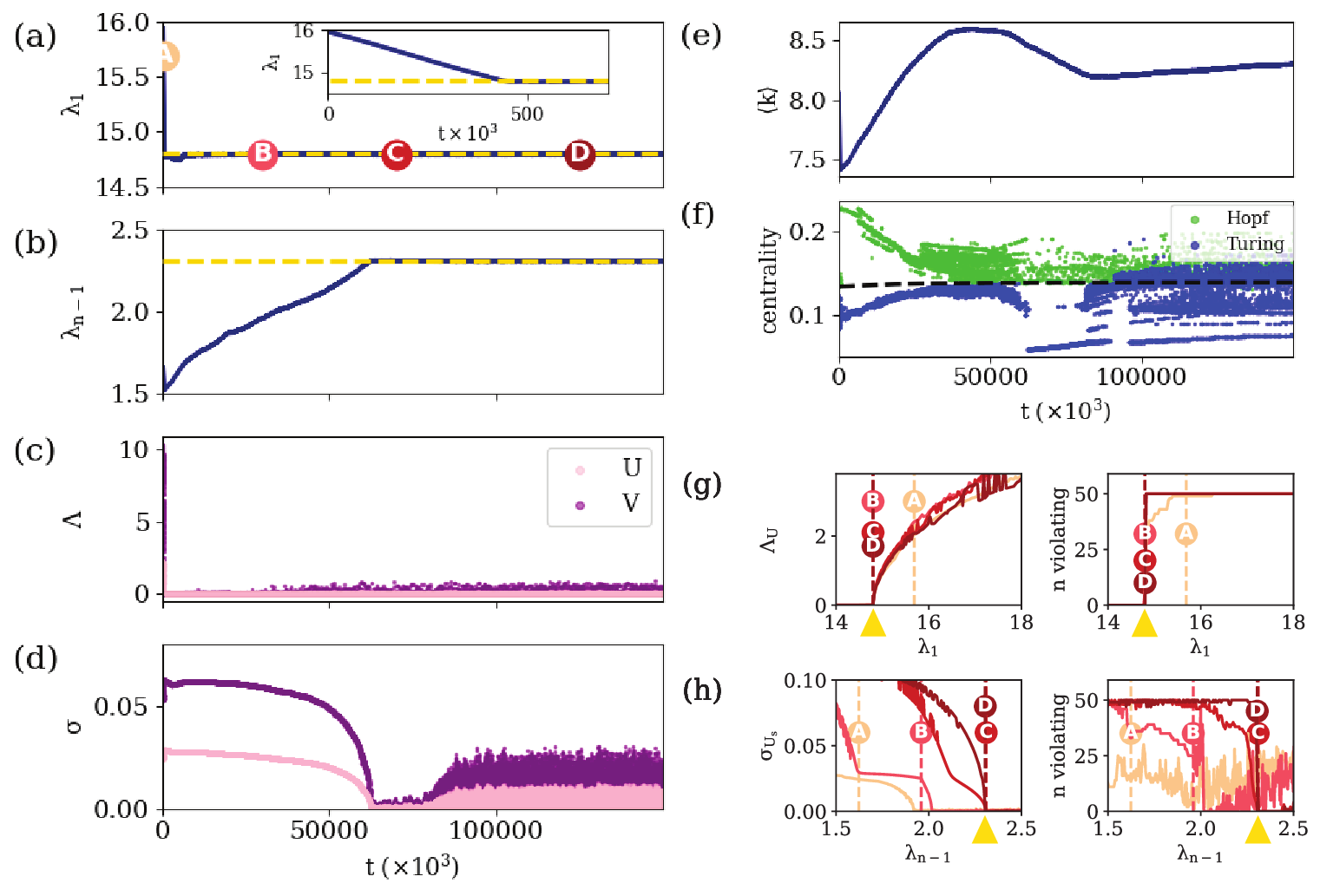}
    \caption{Self-organized drift to multicriticality.  (a-e) Time evolution of the leading Laplacian eigenvalue $\lambda_1$, second smallest Laplacian eigenvalue $\lambda_{n-1}$, Hopf order parameter $\Lambda_U$, Turing order parameter $\sigma_{U_s}$ and the network average degree $\langle k \rangle$. The yellow dashed lines indicate the theoretical critical values $\lambda_{H}^*$ and $\lambda_{T}^*$. The system starts from a supercritical regime with respect to both bifurcations ($\lambda_1 > \lambda_H^*, \lambda_{n-1} < \lambda_T^*$). The parameters of the adaptive mechanisms are set to $\delta w = 0.001, \beta=0.1, \theta_H = 0.05, \theta_T = 0.005, c=10, s=5000$ and $w_{\min} = 0.0001$. (e) 
    Average eigenvector centrality values of nodes whose edge weights are updated in the Hopf or the Turing subcriticality updates.
    (g-h) 
    We freeze the network at time points indicated in panel (a). After turning the adaptive mechanisms off, we create several copies of the frozen network in which we manipulate $\lambda_1$/$\lambda_{n-1}$  by increasing or decreasing all edge weights. We run the simulation on these modified networks with adaptive mechanisms turned off and record the order parameters  $\Lambda_U$ and $\sigma_{U_s}$ to obtain phase diagrams for the Hopf and Turing transitions (left). The dashed lines indicate the actual value of $\lambda_1$/$\lambda_{n-1}$ at the time point in question, while the yellow triangles indicate the critical values $\lambda_H^*$ and $\lambda_T^*$. Note that for time points A and B, the onset of pattern formation does not occur at $\lambda_{n-1} = \lambda^*_T$ because  the modified system is supercritical with respect to the Hopf bifurcation at that point, affecting the bifurcation condition. At time C, the system has reached the Turing bifurcation and the transition occurs at $\lambda_{n-1}= \lambda^*_T$, but the steepness of the phase diagram continues to increase until stabilizing to the form showed for point D. This increase is reflected in the average order parameter $\sigma$ increasing in panel (d) from the early to later stages of the multicritical drift. 
    The right panel shows the number of nodes violating the condition of the Hopf/Turing subcriticality rule, \emph{i.e.}~the number of nodes concluding that their state is indicative of supercriticality.  } 
    \label{fig:timeseries}
\end{figure*}

\section{Results}
To demonstrate multicriticality, we examine how the system evolves when the adaptive mechanisms are let to freely modify the network topology. To start the simulation, we create a random network (Erd\H{o}s-R\'{e}nyi) of size $N=50$ and an average degree of 8 with edge weights set to 1. To induce variation in the edge weights, we add to all edge weights random uniform noise of maximum amplitude 0.1. We initialize the system by setting $(U_i, V_i) = (0.01,0.01)$ for all nodes $i$, after which we let the system evolve on its own.

We observe that as the adaptive mechanisms modify the network topology, the leading Laplacian eigenvalue $\lambda_1$ starts to decrease and soon settles to the critical value $\lambda_H^*$ of the Hopf bifurcation (Fig.~\ref{fig:timeseries}a), indicating that the system has self-organized to the Hopf bifurcation. At this stage, the second smallest Laplacian eigenvalue $\lambda_{n-1}$ lies well below the theoretical critical value $\lambda^*_{T}$ of the Turing bifurcation (Fig.~\ref{fig:timeseries}b). However, as the simulation progresses, while $\lambda_{1}$ remains at the critical value $\lambda_H^*$, $\lambda_{n-1}$ increases to the critical value $\lambda^*_{T}$, indicating that the system has reached multicriticality. In other words, the system has drifted along the Hopf critical manifold to a region of the parameter space where the Turing bifurcation occurs simultaneously, thus becoming multicritical. For the rest of the simulation, $\lambda_{1}$ and $\lambda_{n-1}$ stay at their critical values while other network parameters such as the average weighted degree keep gradually changing (Fig.~\ref{fig:timeseries}e).

To further verify that the system indeed reaches multicriticality, we track how the order parameters of the bifurcations change during the simulation. Order parameter captures the qualitative change characterizing a bifurcation; the Hopf order parameter reflects the presence of oscillations while the Turing order parameter captures pattern formation across nodes. Before each topology update, we analyze the timeseries of $U$ for each node over the last 200 time units, calculating the difference between the largest and the smallest value of $U$, which we define as the oscillation amplitude. The Hopf order parameter $\Lambda_U$ is then the maximum oscillation amplitude across nodes, while the Turing order parameter $\sigma_U$ is the standard deviation of the time-averaged values of $U$. Analyzing the time evolution of these order parameters, we observe that, as expected, the order parameters start to hover close to zero at the same time as the Laplacian eigenvalues indicate that the system has reached the corresponding bifurcation (Figs.~\ref{fig:timeseries}c-d). 

To further validate our observation of multicriticality, we construct phase diagrams of the order parameters at different stages of the simulation to determine the system's distance to the bifurcations 
(Figs.~\ref{fig:timeseries}g-h). We do this by first creating several copies of the network at time $t$  in which we manipulate $\lambda_1/\lambda_{n-1}$ by changing all edge weights by a small amount. After this, we inactivate the adaptive mechanisms and simulate the system for $5000$ time units to measure the order parameters $\Lambda_U$ and $\sigma_U$. The results tell the same story as the Laplacian eigenvalues; 
the system sits at the Hopf/Turing bifurcation when the corresponding eigenvalue $\lambda_1/\lambda_{n-1}$ in Figs.~\ref{fig:timeseries}a-b sits at its theoretical critical value.

For the drift along the critical Hopf manifold to occur, the Hopf and Turing subcriticality rules need to affect $\lambda_1$ and $\lambda_{n-1}$ differently enough. Denoting the effect of the Hopf subcriticality rule on $\lambda_1$ with $\Delta \lambda_{1,H_{sub}}$, it needs to be the case that
\begin{equation}
\frac{\Delta \lambda_{n-1,T_{sub}}}{\Delta \lambda_{1,T_{sub}}} > \frac{|\Delta \lambda_{n-1,H_{sub}}|}{|\Delta \lambda_{1,H_{sub}}|}.
\label{eq:condition}
\end{equation}

\noindent This condition ensures that $\lambda_{n-1}$ is able to increase slowly while $\lambda_1$ stays at its critical value due to adequate time-scale separation between Hopf and Turing updates. 
The condition is met because of the Hopf mechanism targeting the stronger and Turing the weaker links but also because the subcriticality rules tend to operate on nodes with different characteristics. The nodes that first start to oscillate typically have higher-than-average eigenvector centrality values (Fig.~\ref{fig:timeseries}f), which causes the Hopf subcriticality updates to target especially $\lambda_1$. In contrast, the nodes that first start to deviate from their neighbors as the system becomes supercritical with respect to the Turing bifurcation tend to have smaller-than-average eigenvector centralities. Hence, the Hopf subcriticality rule primarily operates on the well-connected nodes, while the Turing rule tends to modify the weakly connected parts of the network. This separation ensures that the adaptive mechanisms operate on $\lambda_1$ and $\lambda_{n-1}$ independently enough to allow the system to drift on the critical Hopf manifold from a region where no Turing bifurcation occurs to a region where it co-occurs. We note that whether or not the condition in Eq.~\ref{eq:condition} is fulfilled depends largely on the network structure; some network realizations fail to reach multicriticality altogether within the simulated time window if the topology evolves in a way that the adaptation mechanisms do not affect the eigenvalues differently enough (see SI VI for an example).

In general, the drift to multicriticality can look significantly different depending on the chosen adaptation parameters (see SI V for analysis on their effect) and the initial network topology. Analyzing the effect of supercriticality updates, we observe that their combined effect on the eigenvalues varies significantly depending on the network structure. In ER networks, the combined effect is to increase $\lambda_1$ and decrease $\lambda_{n-1}$ as expected, but the effects tend to become the opposite for most of the drift before multicriticality is reached (see SI IV). Hence, our model requires the system to be initialized in a supercritical state with respect to both bifurcations, as it is the subcriticality rules rather than the supercriticality ones that originally drive the system to multicriticality. To keep the system at multicriticality, however, the supercriticality rules need to push the eigenvalues to the expected directions. For some network realizations, this happens only after the system has resided at multicriticality for some time, and so the system might drift on an excursion to subcriticality before settling back to multicriticality. Excursions to subcriticality can happen  with respect to one or both of the transitions (see SI VI for examples). We note, however, that while the exact route to multicriticality may vary, the general phenomenon is robust and unaffected by even a substantial variation in the adaptive rules (see SI VII for an example where all edge weights of a node are scaled in each update).

\section{Discussion}
In this work, we have demonstrated for the first time an adaptive system self-organizing to multicriticality. As hypothesized in \cite{gross2021not} and \cite{critical_drift}, viewing criticality not as a point but as a manifold offers a new avenue for exploring complex dynamical behaviors within criticality. Here, we studied a system with two intersecting critical manifolds and showed that two adaptive mechanisms drive the system first to one manifold and then along this manifold to the manifold intersection, where the system becomes critical in two different ways at once. To our knowledge, this constitutes the first demonstration of self-organization to multicriticality.

Overall, these results deepen our understanding of systems in which multiple different bifurcations may be relevant. By demonstrating the inherent flexibility of the criticality hypothesis, our work provides a unifying framework for exploring multicritical manifold intersections and the mechanisms driving systems to these dynamically rich regimes. 
Studying, for instance, the implications of multicriticality on computational performance can lead to new hypotheses on how a system optimizes its behavior in different environments, ultimately resulting in more precise empirical predictions.

In this initial demonstration, we studied a model that allowed analytical treatment with the master stability function approach. This also meant that the occurrence of the bifurcations depended entirely on the Laplacian eigenvalues, which remained constant throughout the multicritical drift. In future work, it would be interesting to study less symmetric models in which even the critical values might change along the drift, and to extend our work by
demonstrating higher-order self-organized multicriticality in more complex models.
 
In general, we expect self-organization to multicriticality to be widespread
generic behaviour,  which has previously been overlooked due to a focus on overly simplified models. If a system can undergo two different bifurcations, the two corresponding critical hypersurfaces will always intersect except in the case where both surfaces are parallel, thus creating a doublecritical hypersurface. If such an intersection exists, self-organization rules that individually bring the system to the respective bifurcations can likely be combined to bring the system to multicriticality. Our initial discovery lays the foundations for exploring the full implications of such dynamical richness.








\begin{acknowledgments}
We acknowledge the computational resources provided by the Aalto University Science-IT project.
\end{acknowledgments}

\InputIfFileExists{bibliography.bbl}

\end{document}